\documentclass[pra,twocolumn,showkeys,showpacs,superscriptaddress,amsmath,amsfonts]{revtex4-1}
\usepackage[latin2]{inputenc}
\usepackage{graphicx}
\usepackage{color}
\def\be{\begin{equation}}
\def\ee{\end{equation}}
\def\bea{\begin{eqnarray}}
\def\eea{\end{eqnarray}}
\def\bi{\begin{itemize}}
\def\ei{\end{itemize}}

\begin{document}

\title{      Overcoming the sign problem at finite temperature:    \\
Quantum Tensor Network for the orbital $e_g$ model on an infinite square lattice}

\author{Piotr Czarnik}
\affiliation{\mbox{Instytut Fizyki im. Mariana Smoluchowskiego, Uniwersytet Jagiello\'nski,
             prof. S. {\L}ojasiewicza 11, PL-30-348 Krak\'ow, Poland}}
\affiliation{\mbox{Institute for Theoretical Physics, Universiteit van Amsterdam,
             Science Park 904, NL-1098 XH Amsterdam, The Netherlands}}

\author{Jacek Dziarmaga}
\affiliation{\mbox{Instytut Fizyki im. Mariana Smoluchowskiego, Uniwersytet Jagiello\'nski,
             prof. S. {\L}ojasiewicza 11, PL-30-348 Krak\'ow, Poland}}

\author{Andrzej M. Ole\'{s} }
\affiliation{\mbox{Instytut Fizyki im. Mariana Smoluchowskiego, Uniwersytet Jagiello\'nski,
             prof. S. {\L}ojasiewicza 11, PL-30-348 Krak\'ow, Poland}}
\affiliation{Max-Planck-Institut f\"ur Festk\"orperforschung,
             Heisenbergstrasse 1, D-70569 Stuttgart, Germany}

\date{\today}

\begin{abstract}
The variational tensor network renormalization approach to
two-dimensional (2D) quantum systems at finite temperature is applied
for the first time to a model suffering the notorious quantum Monte
Carlo sign problem --- the orbital $e_g$ model with spatially highly
anisotropic orbital interactions.
Coarse-graining of the tensor network along the inverse temperature
$\beta$ yields a numerically tractable 2D tensor network representing
the Gibbs state. Its bond dimension $D$ --- limiting the amount of
entanglement --- is a natural refinement parameter. Increasing $D$ we
obtain a converged order parameter and its linear susceptibility close
to the critical point. They
confirm the existence of finite order parameter below the critical
temperature $T_c$, provide a numerically exact estimate of~$T_c$, and
give the critical exponents within $1\%$ of the 2D Ising universality
class.\\
{[\textit{Published in:} Physical Review B \textbf{96}, 014420 (2017)]}
\end{abstract}

\maketitle


\section{Introduction}

Frustration in quantum spin systems occurs by competing exchange
interactions and often leads to disordered spin liquids
\cite{Bal10,Lucile}. This is in contrast to Ising spins on a square
lattice where periodically distributed partial frustration in form of 
exchange interactions with different signs does not suppress a phase 
transition at finite temperature $T_c$ \cite{Lon80}, while complete
frustration gives a disordered classical phase \cite{Vil77}.
Frustration may also be generated by a different mechanism --- when
Ising-like interactions for different pseudospin components compete on
a square lattice in the two-dimensional (2D) compass model
\cite{Nus05,Dou05,Dor05,Tro10} or on the honeycomb lattice in the
Kitaev model \cite{Kit06}. While the short-range spin liquid is
realized in the Kitaev model \cite{Bas07}, the pseudospin nematic
order stabilizes below $T_c$ in the 2D compass model \cite{Wen10,Cza16}.
In such cases entanglement plays an important role \cite{Ami08}
and advanced methods of quantum many-body theory have to be applied.

In real systems pseudospin interactions concern the orbital degrees of
freedom. The case of $e_g$ orbitals is paradigmatic here as it (i) is
related to the 2D compass model \cite{Cin10} and (ii)
initiated spin-orbital physics \cite{Kug82,Fei97,Tok00,Cor12,Brz15}
--- the well known systems with $e_g$ orbitals are:
KCuF$_3$ \cite{Bella,Dei08,Pav08},
LaMnO$_3$ \cite{Dag01,Fei99,Ish02,Ish03,Ole05,Pav10,Kov10,Sna16}, and
LiNiO$_2$ \cite{Rey01,Mil04,Rei05}.
This field is very challenging due to the interplay and entanglement
of spins and orbitals which leads to remarkable consequences
\cite{Ole12,Brz12}. However, when spin order is ferromagnetic,
as in the $(a,b)$ planes of KCuF$_3$ and LaMnO$_3$, spins disentangle
and one is left with the 2D orbital $e_g$ model \cite{vdB99,Tan05}
where hole propagation is possible by the coupling to orbitons
\cite{vdB00}. Surprisingly, the tendency towards long-range order with
such excitations is then \textit{opposite} to that for spin systems
\cite{Mat06}, i.e., $e_g$ orbital order occurs in a 2D square lattice
below $T_c$ \cite{Ryn10,Wen11}, for instance in K$_2$CuF$_4$
\cite{kcuf4,Mos04}, while the role of quantum fluctuations increases
with increasing dimension \cite{vdB99,Fei05}.

In this article we investigate a phase transition at $T_c$ in the 2D
orbital $e_g$ model.
A better understanding of the signatures of this phase transition
provides a theoretical challenge. We present a very accurate estimate
of $T_c$ and the critical exponents being in the 2D
Ising universality class. These results could be achieved due to a
remarkable recent progress in tensor networks due to the
formulation of an algorithm at finite temperature using
a projected entangled-pair operator (PEPO) \cite{var}.

The paper is organized as follows. Sec. \ref{sec:TN} gives brief 
overview of tensor network methods. Sec. \ref{sec:mod} introduces 
simulated model. Sec. \ref{sec:alg} introduces 2D finite temperature 
tensor network method used to simulate the model. Numerical results 
are presented in Sec. \ref{sec:res}. Sec. \ref{sec:summ} summarizes 
the paper. Appendix \ref{sec:conv} gives detailed description of results 
convergence analysis which enabled us to obtain trustworthy results 
for the model. Technical details of simulations are given in Appendix
\ref{sec:num_det}. Finally Appendix \ref{sec:lowT} gives additional  
results for low temperature regime of the model.

\section{Tensor networks}
\label{sec:TN}
Since the discovery of the density matrix renormalization group (DMRG)
\cite{White,Sch05} --- that was later shown to optimize the matrix
product state (MPS) variational ansatz \cite{Sch11} --- quantum tensor
networks proved to be an indispensable tool to study strongly correlated
quantum systems \cite{Che15}. MPS ansatz was later generalized to a 2D
projected entangled pair state (PEPS) \cite{PEPS,CMRPEPS} and supplemented with
the multiscale entanglement renormalization ansatz (MERA) \cite{MERA}. The
networks do not suffer from the notorious sign problem \cite{fermions}
and in the doped case fermionic PEPS provided better variational energies
for the $t$-$J$ model \cite{PEPStJ} and the Hubbard model \cite{Cor15}
than the best available variational Monte Carlo results. A combination of
different tensor networks, supplemented with other sign-error free
methods, seems to have finally settled the controversy on the ground
state of the underdoped Hubbard model \cite{stripesHubbard}.
The networks --- both MPS \cite{Yan11,Cin13,Ran15} and PEPS
\cite{Poi12,Poi13,Wan13} --- also made some major breakthroughs in the
search for topological order. This is where, like in the $e_g$ model
\cite{Ryn10}, geometric frustration often prohibits the traditional
quantum Monte Carlo.

Thermal states of quantum Hamiltonians were explored much less than
their ground states. In one dimension they can be represented by an
MPS ansatz prepared with an accurate imaginary time evolution
\cite{Ver04,WhiteT}. A~similar approach can be applied in 2D models
\cite{Czarnik,self}, where the PEPS manifold is a compact
representation for Gibbs states \cite{Molnar} but the accurate
evolution proved to be more challenging. Alternative direct
contractions of the 3D partition function were proposed \cite{ChinaT}
but, due to local tensor update, they are expected to converge more
slowly with increasing refinement parameter. Even a small improvement
towards a full update can accelerate the convergence significantly
\cite{HOSRG}.

In order to avoid these problems, in the pioneering work \cite{var} two
of us introduced an algorithm to optimize variationally a projected
entangled-pair operator (PEPO) representing the Gibbs state
$e^{-\beta H}$ of a 2D lattice system ($\beta\equiv 1/T$). Its first
challenging benchmark applications include the quantum compass
\cite{Cza16} and Hubbard \cite{varHubbard} models where it provided
accuracy comparable to the best conventional methods.

It was not quite unexpected. Just like for the ground-state PEPS, the
accuracy of the thermal PEPO is limited by its finite bond dimension
$D$, i.e., the size of tensor indices connecting nearest-neighbor
lattice sites. This size limits the entanglement within the
ground/thermal state. However, by its very definition the Gibbs state
is the mixed state that maximizes the entropy for a given average
energy. Since this maximal entropy is actually the entropy of
entanglement with the rest of the universe, then --- thanks to the
monogamy of entanglement --- the Gibbs state also minimizes its
internal entanglement. Among all states with the same average energy
it is the one most suited to be represented by a tensor network.
Encouraged by the benchmarks tests, in this work we apply the algorithm
for the first time to a model that evades treatment by quantum Monte
Carlo \cite{Ryn10,Wen11}. Numerical convergence and
self-consistency alone allow us to make definitive statements on the
physics of the model demonstrating the power of this method.

\begin{figure}[t!]
\vspace{-0cm}
\includegraphics[width=0.88\columnwidth,clip=true]{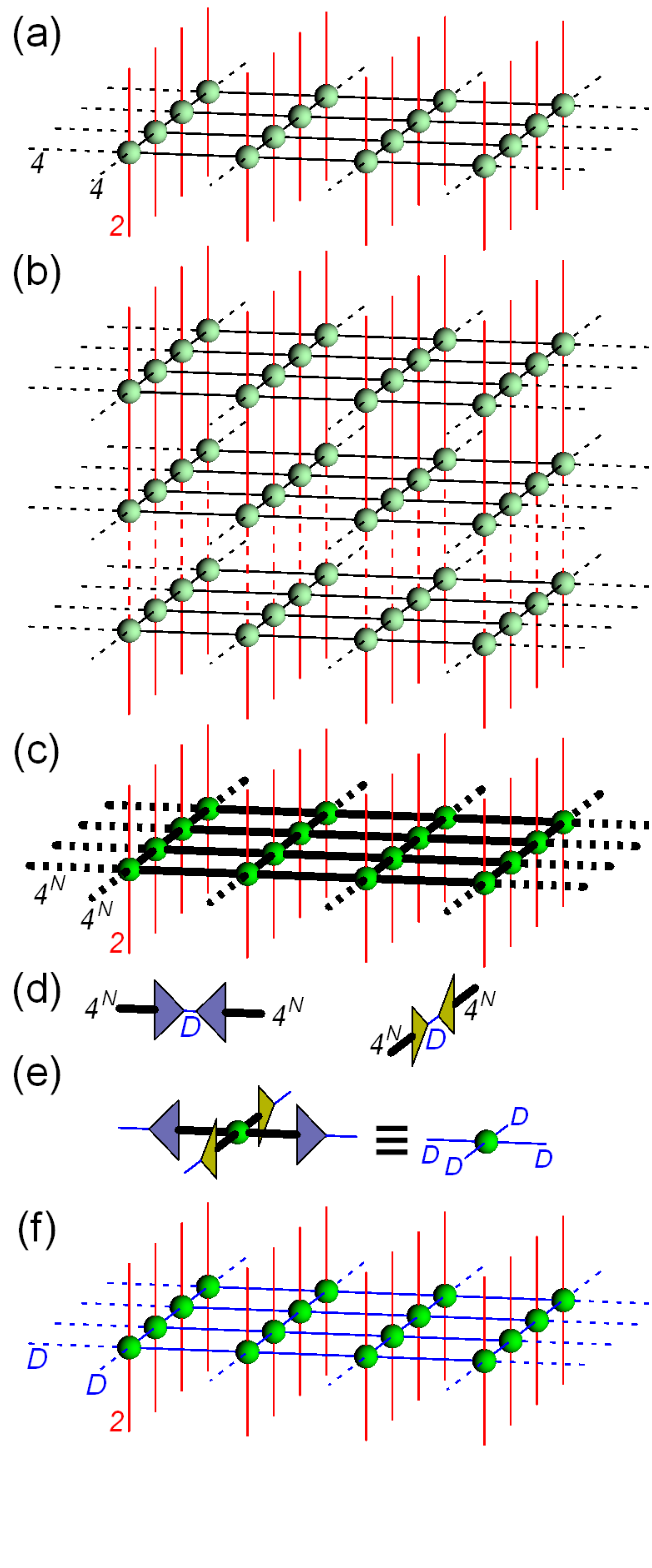}
\vspace{-0.5cm}
\caption{
A route towards a tractable 2D PEPO network:
(a)~a~small time step $U(d\beta)$ as a PEPO network with a bond
dimension $4$;
(b) the operator $e^{-\beta H/2}\equiv U(\beta)$ as a product of
$N$ small steps $U(d\beta)^N$ ---
contraction of (b) along each column gives
(c) a 2D network with a huge bond dimension $4^N$ where
each bond line 
is inserted with (d) an orthogonal projection of
dimension $D$ made of two isometries; next
each isometry is absorbed into its (e) nearest tensor truncating the
dimension of its bond index from $4^N$ down to $D$. It leads to
a~network $U(\beta)$ depicted in (f) with a bond dimension $D$.
}
\label{fig:3Dto2D}
\end{figure}

\section{The $e_g$ orbital model}
\label{sec:mod}

The quantum $e_g$ model on an infinite square lattice is defined by
the Hamiltonian
\begin{equation}
H = - J\sum_j \sum_{\alpha=a,b} \tau^{\alpha}_j \tau^{\alpha}_{j+e_\alpha}.
\label{eg}
\end{equation}
Here $j$ labels lattice sites, $e_a(e_b)$ are unit vectors along the
$a(b)$ axis and $\tau^\alpha_j$ are orbital operators represented by
Pauli matrices:
\bea
\tau^a_j = \frac14\left( -\sigma^z_j + \sqrt3 \sigma^x_j \right),~~
\tau^b_j = \frac14\left( -\sigma^z_j - \sqrt3 \sigma^x_j \right).
\eea
The coupling in the orbital space depends on the spatial orientation
of the bond. In what follows $J=1$.

At low temperature a spontaneous breaking of symmetry takes place and
the system orders according to the strongest interaction
$\propto\frac{3}{16}\sigma^x_i\sigma^x_j$ \cite{Cin10}.
This symmetry breaking implies a finite real order parameter
\be
m(T)\equiv\langle\sigma^x_j\rangle.
\label{m}
\ee
Unlike the 2D compass model \cite{Wen10},
the model (\ref{eg}) is not tractable by Monte Carlo \cite{Wen11},
but the order parameter suggests the 2D Ising universality class for
the finite temperature transition which is confirmed
by our simulations.

\section{The algorithm at $T>0$}
\label{sec:alg}
The algorithm was described in all technical detail elsewhere
\cite{Cza16}. Its aim is to represent matrix elements of the operator
$\rho=e^{-\beta H/2}$ by the 2D tensor network in Fig.
\ref{fig:3Dto2D}.
Here we show only a small $4\times 4$ unit
of an infinite square lattice and each geometrical shape
(here a green ball) represents a tensor. There is one tensor at every
lattice site. Each line sticking out of the tensor represents one index.
A (black) line connecting two tensors represents a tensor contraction
through the connecting index. There is one bond index along every
nearest neighbor bond. It has a finite bond dimension $D$. The dashed
bond lines connect the $4\times 4$ unit with the rest of the
lattice. The open (red) vertical indices number the orbital basis'
states. Those pointing up/down number bra/ket states. The desired 2D
network in Fig.~\ref{fig:3Dto2D}(f) --- known as PEPO --- can be
contracted efficiently to obtain local expectation values. A finite $D$
is sufficient to represent Gibbs states with their limited entanglement.

On the other hand, the 2D operator $e^{-\beta H/2}\equiv U(\beta)$ can
be naturally represented by a 3D network, the third dimension being
the imaginary time $\beta$. The evolution is split into $N$ small time 
steps ($d\beta\ll 1$), $U(\beta)=U(d\beta)^N$. With a
Suzuki-Trotter decomposition, each step can be represented by a 2D
layer in Fig.~\ref{fig:3Dto2D}(a). In the $e_g$ model, its bond indices
have dimension $4$. The product of $N$ steps is the 3D network in Fig.
\ref{fig:3Dto2D}(b). Here we show only three layers;
the remaining $N-3$ ones are represented by the vertical dashed lines.

The 3D network is too hard to treat directly. Formally, it can be
compressed to a 2D network by contracting along each vertical column
first. The resulting 2D network in Fig.~\ref{fig:3Dto2D}(c) arises
at the price of a huge
bond dimension $4^N$. Fortunately, we know that just a tiny
$D$-dimensional subspace in the $4^N$ dimensions is enough to
accommodate all correlations. Therefore, it is justified to insert every
bond line with a $D$-dimensional projection made of two isometries.
There are two independent projections along the axes $a$ and $b$,
see Fig.~\ref{fig:3Dto2D}(d). After the insertion, every isometry is
absorbed into its nearest tensor truncating its bond index down to a
tractable size $D$, see Fig.~\ref{fig:3Dto2D}(e). The outcome is the
desired PEPO $U(\beta)$ in Fig.~\ref{fig:3Dto2D}(f), and the Gibbs
state is $e^{-\beta H}=U^\dag(\beta)U(\beta)$.

Now the problem is how to handle the huge isometries from $4^N$ to $D$.
Fortunately,
by a divide-and-conquer strategy, each of them can be
split into a hierarchy of smaller isometries connected into a tree
tensor network \cite{Cza16}. It is possible to optimize the smaller
isometries one-by-one to obtain the most accurate projection available
for a given $D$. The cost of the algorithm is polynomial in $D$ and
only logarithmic in the number of steps $N$, allowing for $d\beta$
small enough to make the Suzuki-Trotter decomposition numerically
exact at very little expense.

\begin{figure}[t!]
\vspace{-0cm}
\includegraphics[width=0.9\columnwidth,clip=true]{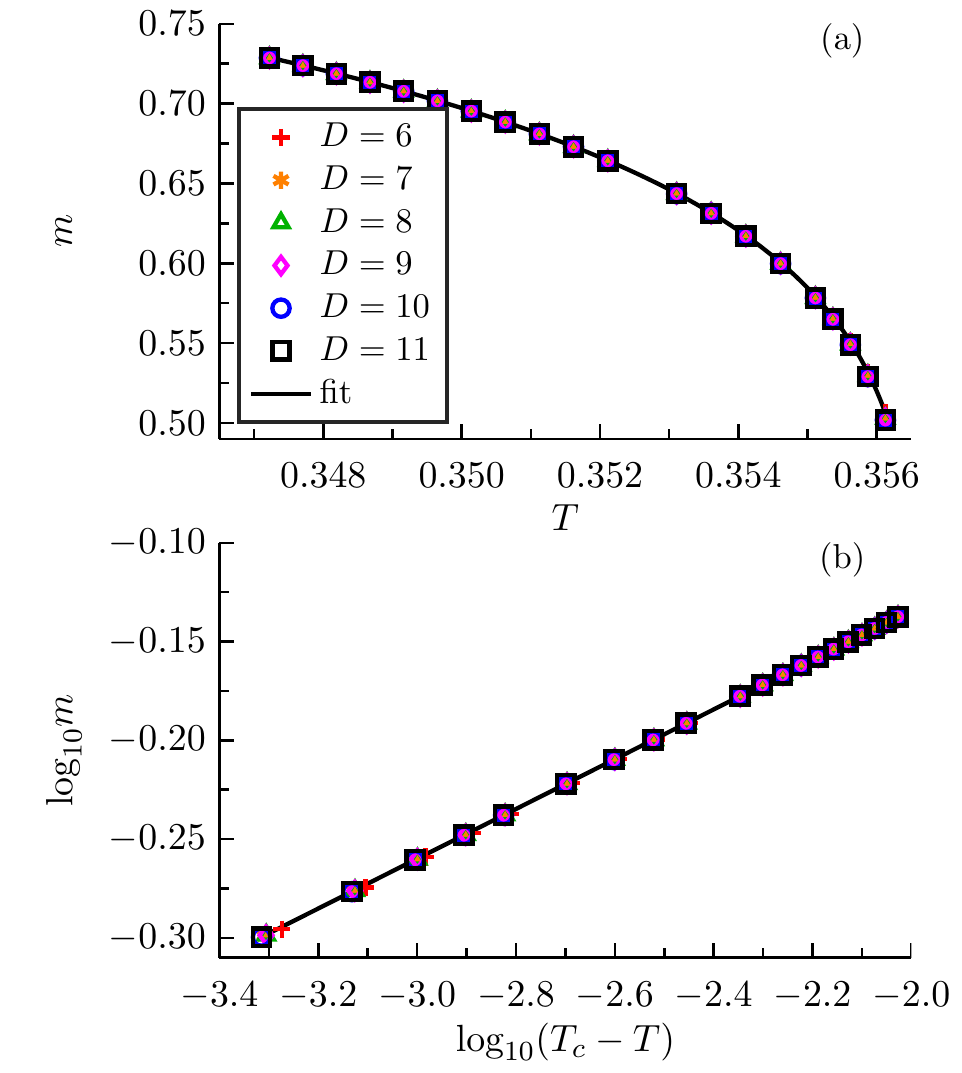}
\vspace{-0cm}
\caption{
The order parameter $m=\langle\sigma^x\rangle$ (\ref{m}) for increasing
temperature $T$ for different bond dimensions $D$. The solid line is the
best fit
in Eq. (\ref{mT}) to the results for $D=11$. Figure \ref{fig:convD}
demonstrates that they are already converged in $D$.
}
\label{fig:plotm}
\end{figure}

\section{Numerical Results}
\label{sec:res}
\begin{figure}[t!]
\vspace{-0cm}
\includegraphics[width=0.91\columnwidth,clip=true]{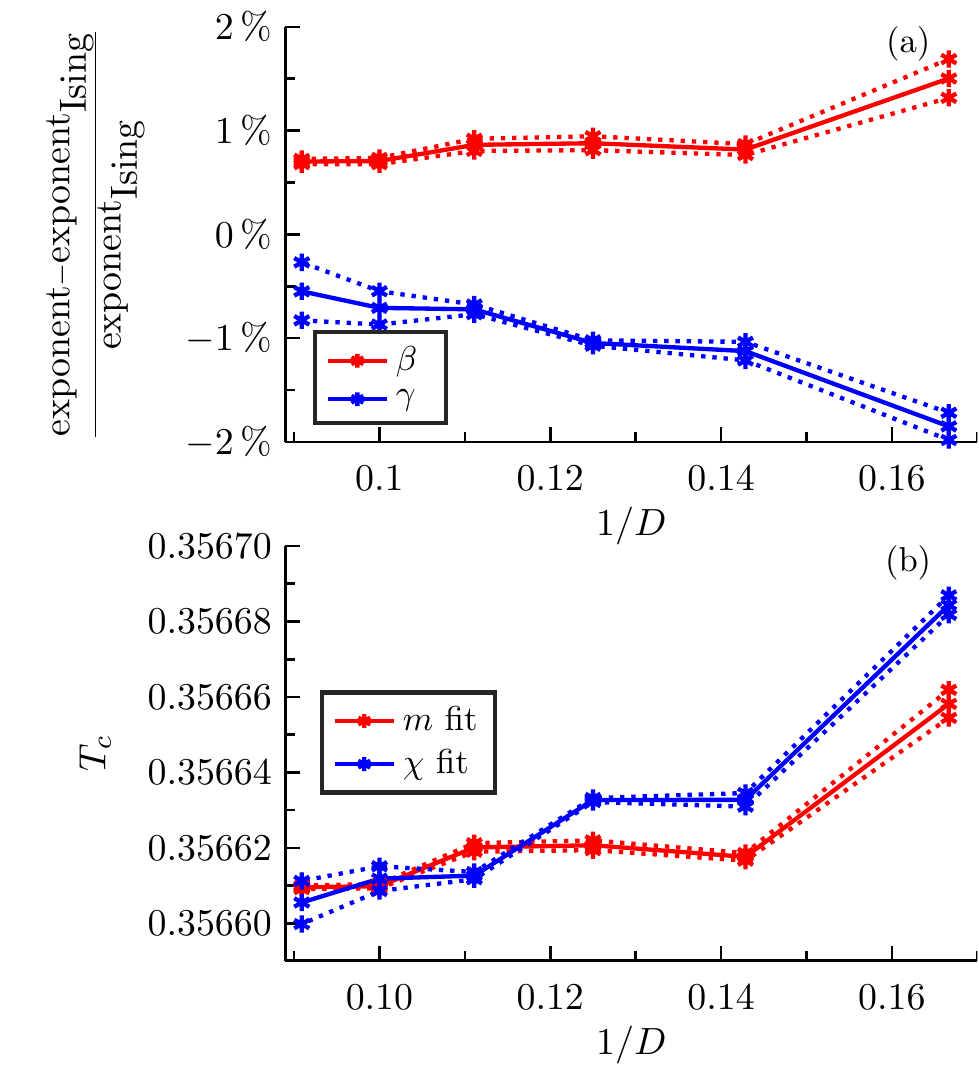}
\vspace{-0cm}
\caption{
Convergence tests as functions of the inverse bond dimension $1/D$:
(a) the relative differences between the fitted critical exponents
$\{\beta,\gamma\}$ and their 2D Ising values $\frac18$ and $\frac74$;
(b) the fitted critical temperatures $T_c$ from
$m(T)$ in Fig.~\ref{fig:plotm} and
$\chi(T)$ in Fig.~\ref{fig:plotsus}.
The solid lines connect
the best fits and the dashed lines delimit their error bars.
For the selected $T$
intervals close to $T_c$
fitted results depend primarily on $D$, see Appendix \ref{sec:conv}, 
Figs. \ref{fig:beta_vs_interval} and \ref{fig:gamma_vs_interval}.
}
\label{fig:convD}
\end{figure}

For each $T<T_c$
the order parameter $m$ (\ref{m}) was converged in $D$ in the symmetry
broken phase,
see Fig.~\ref{fig:plotm}. For each $D$ it was fitted with a power law,
\be
m(T)\propto(T_c-T)^{\beta},
\label{mT}
\ee
see Fig.~\ref{fig:convD}. Here $\beta$
is the order-parameter critical exponent (not to be confused with the
inverse temperature $\beta=1/T$). For $D\ge 7$ the estimates:
$0.35660<T_c< 0.35664$, and
$0.1258 <\beta< 0.1261$, do not depend significantly on increasing $D$.
They slowly drift towards $T_c=0.35661$ and $\beta=0.125$, respectively.
For more details see Appendix \ref{sec:conv}.

In the symmetric phase above $T_c$, we calculated the magnetic
susceptibility using the linear approximation,
\be
\chi(T)=\left.\frac{dm}{dh}\right|_{h=0}.
\label{chiA0}
\ee
Here $h$ is an infinitesimal symmetry-breaking field $h\sum_i\tau_i^x$
added to the Hamiltonian (\ref{eg}). The derivative was approximated
accurately by a finite difference between $h=10^{-6}$ and $h=0$.
More details on $\chi(T)$ numerical calculation are given in  
Appendix \ref{sec:num_det},
see Fig.~\ref{fig:susanis} and Table~\ref{tab:deltah}.

\begin{figure}[t!]
\vspace{-0cm}
\includegraphics[width=0.91\columnwidth,clip=true]{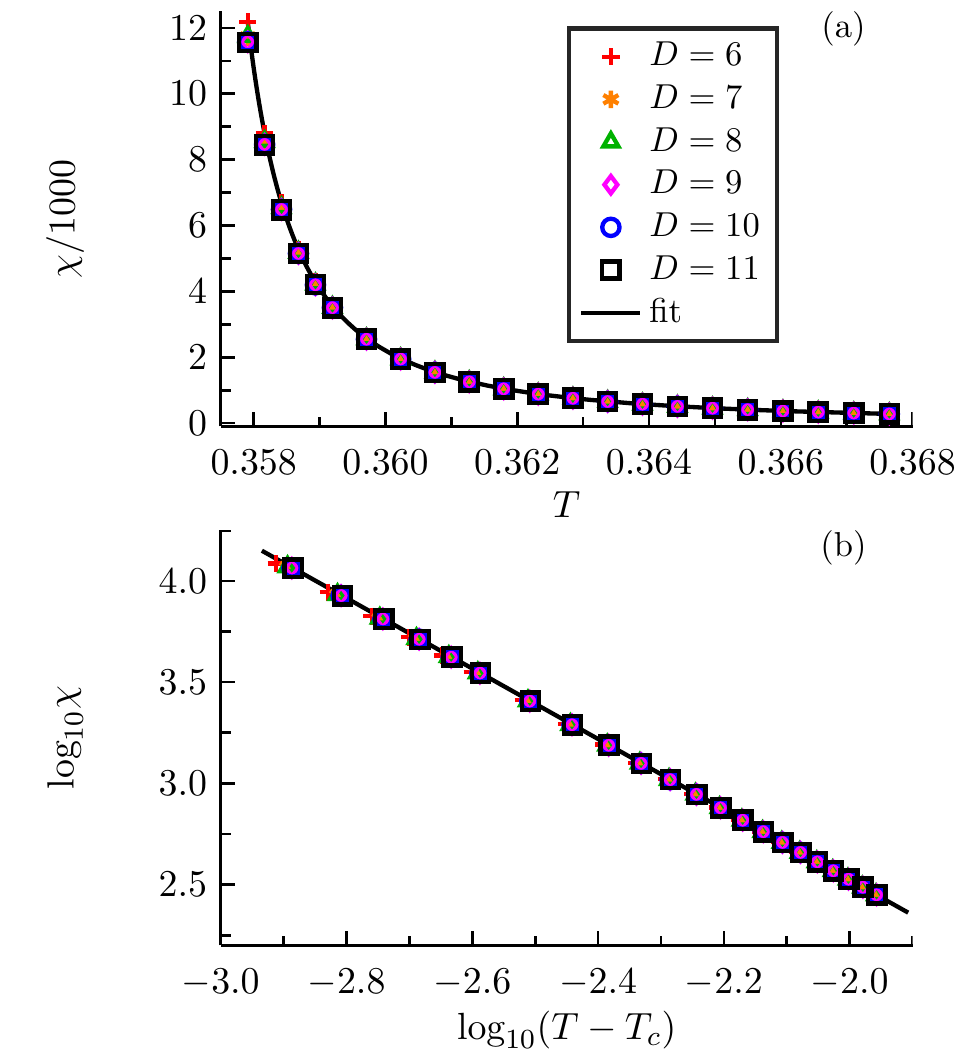}
\vspace{-0cm}
\caption{
The linear susceptibility $\chi(T)$ in the symmetric phase
(\ref{chiA0}) for different bond dimension $D$.
The solid line is the best fit of the power law (\ref{chiT}) to
the results for $D=11$. Figure \ref{fig:convD} demonstrates that
they are already converged in $D$.
}
\label{fig:plotsus}
\end{figure}

The susceptibility was converged in $D$ (Fig.~\ref{fig:plotsus})
and fitted with a power law,
\be
\chi(T)\propto(T-T_c)^\gamma,
\label{chiT}
\ee
see Fig.~\ref{fig:convD} and Appendix \ref{sec:conv}.
Again, for $D\ge 7$ the estimates:
$0.35660<T_c<0.35665$, and $1.732<\gamma<1.740$, almost
do not depend on increasing $D$, and drift towards
$T_c=0.35661$ and $\gamma\simeq 1.75$.
Altogether, both exponents are less than $1\%$ away from the exact 
$\beta=\frac18$ [see Fig. 5(a)] and $\gamma=\frac74$
in the 2D Ising universality class.

Remarkably, $T_c$ found from $m(T)$ (\ref{mT}) and $\chi(T)$
(\ref{chiT}) is \textit{identical} up to the four-digit precision.
We propose
\be
T_c = 0.3566 \pm 0.0001,
\label{Tc}
\ee
deduced from the scatter of the data for $D\ge 7$ in Fig.~\ref{fig:convD}(b) 
multiplied by a factor of $3$, see also Fig. \ref{fig:beta_vs_interval}(b). 
It is worthwhile to compare the above estimate (\ref{Tc}) with the 2D 
Ising model \cite{Ons44} with interaction $\frac14\sigma^z_i\sigma^z_j$,
\be
T_c^{\rm Ising} = \frac{1}{2\log(1+\sqrt{2})}\approx 0.567296.
\label{TcIsing}
\ee
Exchange interactions in the dominating term 
$\frac{3}{16}\sigma^x_i\sigma^x_j$ in Eq. (\ref{eg}) are reduced by the 
factor $\frac34$ from the 2D Ising model, so this reduction alone would 
give instead $T_c=0.75 T_c^{\rm Ising}$.
\textit{De facto}, the obtained value in Eq. (\ref{Tc}) is
$T_c\simeq 0.6286 T_c^{\rm Ising}$, i.e., it is further reduced by
$\sim 16$\% by quantum fluctuations activated at finite $T$ due to
$\propto\frac{\sqrt{3}}{4}(\sigma^x_i\sigma^z_j+\sigma^x_i\sigma^z_j)$
and $\propto\frac14\sigma^z_i\sigma^z_j$ terms in Eq. (\ref{eg}).
The order parameter (\ref{m}) at $T=0$ is almost saturated
as quantum fluctuations are negligible at $T\to 0$,
\be
m(0)=0.993.
\ee
More details on $m(0)$ simulation are given in Appendix \ref{sec:lowT}, 
see Fig.~\ref{fig:ordlowT}. The value in Eq. (9) was obtained by the 
present 
method and agrees with the ground state MERA calculations \cite{Cin10}. 
This shows that the quantum fluctuation effects in the $e_g$ orbital 
model (1) are very weak indeed at $T=0$ \cite{vdB99}, while at $T>0$ 
the fluctuations are activated and reduce significantly the value of 
the critical temperature down to $T_c\simeq 0.3566$, see Eq. (\ref{Tc}).
Indeed quantum fluctuations play a role here but are not as significant
as for the 2D SU(2) symmetric Heisenberg antiferromagnet \cite{Mat06}.
Yet, the entanglement between the orbital operators is here much
reduced from that in the 2D compass model \cite{var} and therefore
such an accurate estimate of $T_c$ (\ref{Tc}) is possible.

\section{Summary}
\label{sec:summ}

Being a paradigmatic frustrated system, the orbital
$e_g$ model evades treatment by quantum Monte Carlo but it proves to
be accurately tractable by our thermal tensor network. The notorious
sign problem --- often inescapable for quantum Monte Carlo ---
is not an issue for our method. Instead the relevant issue is if the
entanglement in a thermal state can be accommodated within a bond
dimension that is small enough to fit into a classical computer.
This criterion is satisfied by the thermal state of the $e_g$ model
and a four-digit estimate of the critical temperature and a better
than $1\%$ accuracy of the critical exponents could be achieved.
Since the Gibbs state is the least entangled one among all excited
states with the same average energy, it is potentially the easiest
target for a suitable tensor network.

\acknowledgments
We thank Philippe Corboz for insightful discussions.
We kindly acknowledge support by Narodowe Centrum Nauki
(NCN, National Science Centre, Poland) under Projects:
No.~2013/09/B/ST3/01603 (P.C. and J.D.)
and
No.~2016/23/B/ST3/00839 (A.M.O.).
The work of P.C. on his Ph.D. thesis was supported
by NCN under Project No.~2015/16/T/ST3/00502.


\appendix

\section{Convergence of the results}
\label{sec:conv}

The bond dimension $D$ (see Fig.~1) has to be large enough to
accommodate the entanglement in the thermal state. Furthermore, an
environmental bond dimension $M$ that is used in the analysis of the
effective 2D tensor network depicted in Fig.~1(f)
(see Ref. \cite{Cza16} for details) has to be large enough to 
accommodate long range correlations. In general, these requirements 
cannot be satisfied at the critical temperature $T_c$ but the phase 
transition can be approached from both sides close enough to fit the 
critical power laws. In this appendix we demonstrate that indeed we 
are able to approach $T_c$ close enough to obtain stable and 
converged fits.

All results presented here, which were obtained with $M=72$, are 
converged in $M$. Another potential source of errors are Trotter errors. 
They are not a significant issue for our approach as its cost scales at 
most logarithmically with the the inverse Trotter time step $1/d\beta$. 
Our results were obtained with
$d\beta\le0.001$ and are converged in $d\beta$.

\begin{figure}[t!]
\vspace{-0cm}
\includegraphics[width=0.97\columnwidth,clip=true]{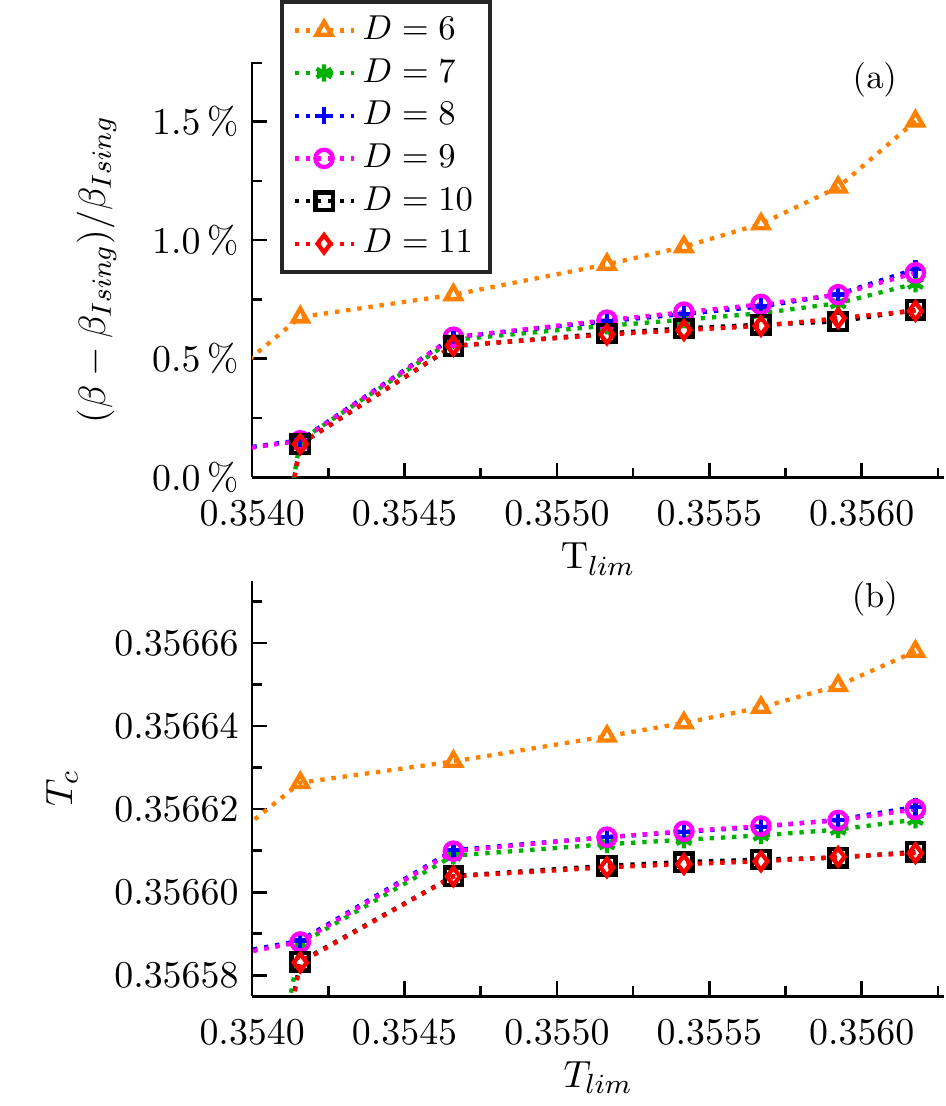}
\vspace{-0cm}
\caption{
The dependence of the
(a) exponent $\beta$ and
(b) critical temperature $T_c$
obtained by fitting $m(T)$ for different $D$ (shown in Fig.~2)
within the range of temperature  $0.3472<T<T_{lim}$.
For $D\geq7$, with increasing $T_{lim}$ approaching the critical point, 
the fitted $T_c$ approaches $T_c = 0.3566$ becoming stable with respect 
to the choice of $T_{lim}$, while the fitted  $\beta$ stabilizes within 
$1\%$ of  $\beta_{\rm Ising}=1/8$
drifting slowly towards $\beta_{\rm Ising}$ with increasing $D$.
}
\label{fig:beta_vs_interval}
\end{figure}

\begin{figure}[t!]
\vspace{-0cm}
\includegraphics[width=0.97\columnwidth,clip=true]{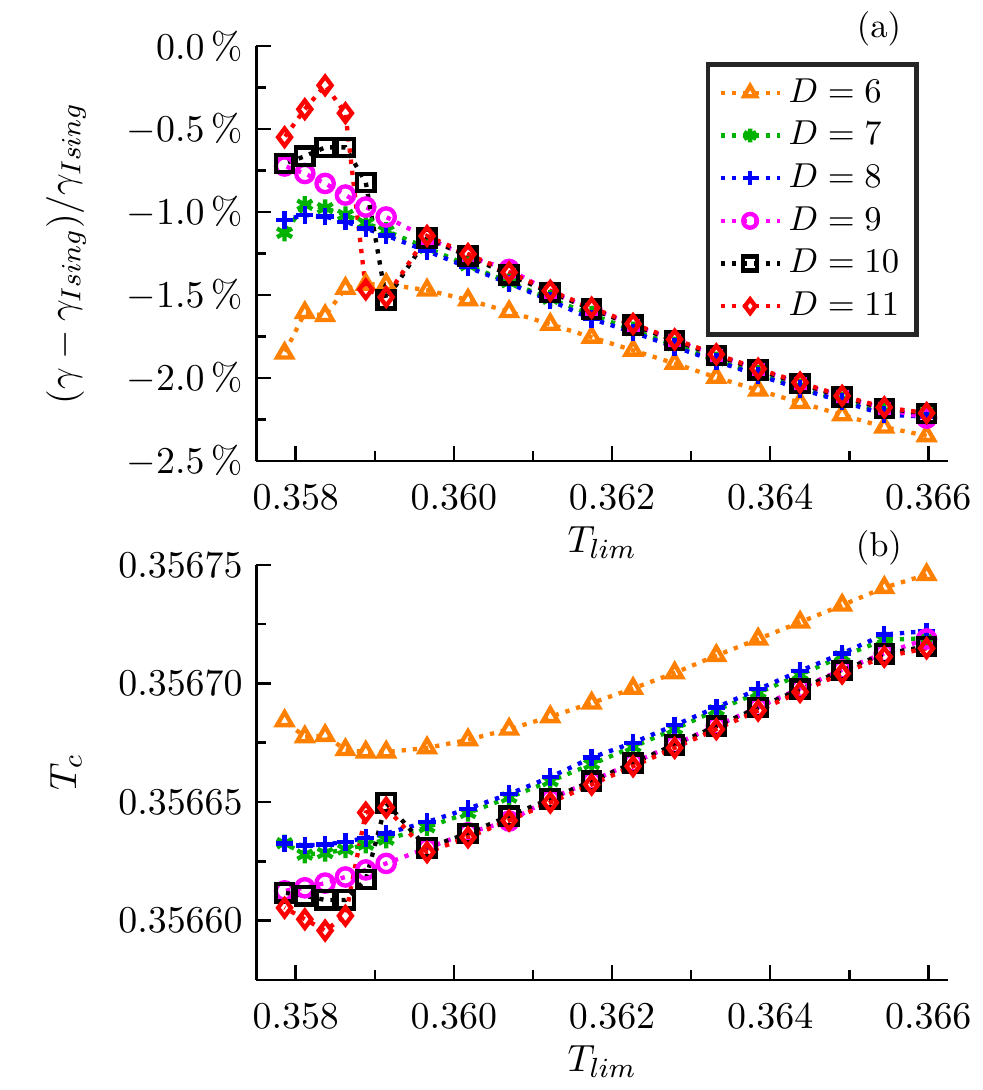}
\vspace{-0cm}
\caption{
The dependence of the
(a) exponent $\gamma$ and
(b) critical temperature $T_c$
obtained by fitting  $\chi(T)$ (shown in Fig.~4)
within the  range of temperature $0.3677>T >T_{lim}$.
For $D\geq7$ with decreasing $T_{lim}$ approaching the critical point
the fitted $T_c$ approaches $T_c = 0.3566$. Close to the smallest
value of $T_{lim}$  it becomes dependent primarily on $D$.  
Similar behavior occurs for $\gamma$ which for  $D\geq7$ approaches
$\gamma_{\rm Ising}$ with decreasing $T_{lim}$ becoming finally  
primarily $D$-dependent and drifting towards $\gamma_{\rm Ising}$
with increasing $D$.  }
\label{fig:gamma_vs_interval}
\end{figure}

The convergence of the critical exponents, $\beta$ for the
magnetization $m(T)$ and $\gamma$ for the susceptibility $\chi(T)$, 
is shown in Figs. \ref{fig:beta_vs_interval}(a) and
\ref{fig:gamma_vs_interval}(a) where we compare them with the
2D Ising model exponents,
\be
\beta_{\rm Ising}=\frac18, \hskip .7cm \gamma_{\rm Ising}=\frac74.
\label{Ising}
\ee
For $D\ge7$ we see that the exponents  approach the Ising values
while  $T_{lim}$ is approaching $T_c$. For $T_{lim}$ sufficiently close
to $T_c$ they no longer depend significantly on range of $T$ depending
instead primarily on $D$. In this regime all fitted exponents fall
within $1 \%$  of 2D Ising universality class, drifting towards
$\beta_{Ising}$ or $\gamma_{\rm Ising}$ with increasing $D$.
The obtained behavior of the exponents indicates the 2D Ising
universality class of the transition.

The data collected in Figs. \ref{fig:beta_vs_interval}(b) and
\ref{fig:gamma_vs_interval}(b) demonstrate similar convergence behavior 
of fitted $T_c$ as for the exponents. For $D\ge7$ fitted $T_c$  
approaches $T_c=0.3566$ when $T_{lim}$ is approaching the critical 
point.  For $T_{lim}$ sufficiently close to $T_c$ the critical point
$T_c$ begins to depend primarily on $D$ rather than on $T_{lim}$.
Reaching this regime where the fits become stable with respect to 
$T_{lim}$ justifies taking into account only their $D$ dependence to 
obtain the final $T_c$ estimate Eq. (\ref{Tc}).

We remark that  our estimate of $T_c$ is based on two independent $T_c$
estimates,  coming either from the $\chi(T)$ or $m(T)$ fits, which agree 
up to five digits for the largest~$D$.

\section{Numerical details}
\label{sec:num_det}

\begin{figure}[b!]
\vspace{-0cm}
\includegraphics[width=\columnwidth,clip=true]{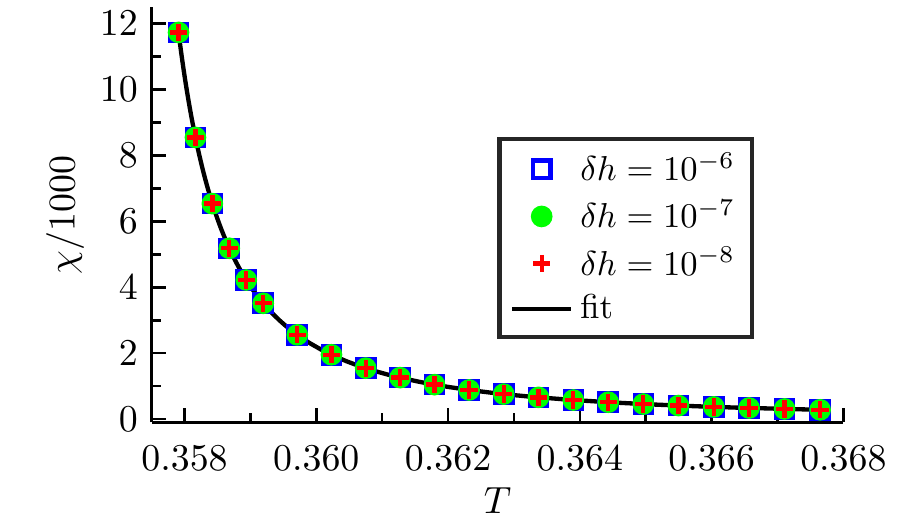}
\vspace{-0.3cm}
\caption{
The linear susceptibility $\chi(T)$ in the symmetric phase
(\ref{chiA0}) obtained  for different symmetry breaking field values 
$\delta h$  with  $D=8$ and $M=72$.
The solid line is the best fit of the power law (\ref{chiT}) to
the results. The figure demonstrates that $\chi(T)$
is already converged in $\delta h$ for $\delta h = 10^{-6}$ used in 
Fig.~\ref{fig:plotsus}.
}
\label{fig:susanis}
\end{figure}

\begin{table}[t!]
\centering
\begin{ruledtabular}
\begin{tabular}{ccccc}
& $\delta h$ &  $T_c$     &   $\gamma$  &   \\ \hline
& $10^{-6}$  & 0.356631   &    1.7324   &   \\
& $10^{-7}$  & 0.356633   &    1.7317   &   \\
& $10^{-8}$  & 0.356633   &    1.7317   &   \\
\end{tabular}
\end{ruledtabular}
\caption{Fitted $T_c$ and $\gamma$ obtained for different symmetry 
breaking field values $\delta h$ with $D=8$ and $M=72$. Here data for  
$0.3566 < T < 0.3677$ were used. Changes of the fitted $T_c$ and 
$\gamma$ with decreasing $\delta h\le 10^{-6}$ are negligible as 
compared to their dependence on $D$ or range of data used to fit 
$T_c$ and~$\gamma$.
}
\label{tab:deltah}
\end{table}

\begin{figure}[b!]
\vspace{-0cm}
\includegraphics[width=\columnwidth,clip=true]{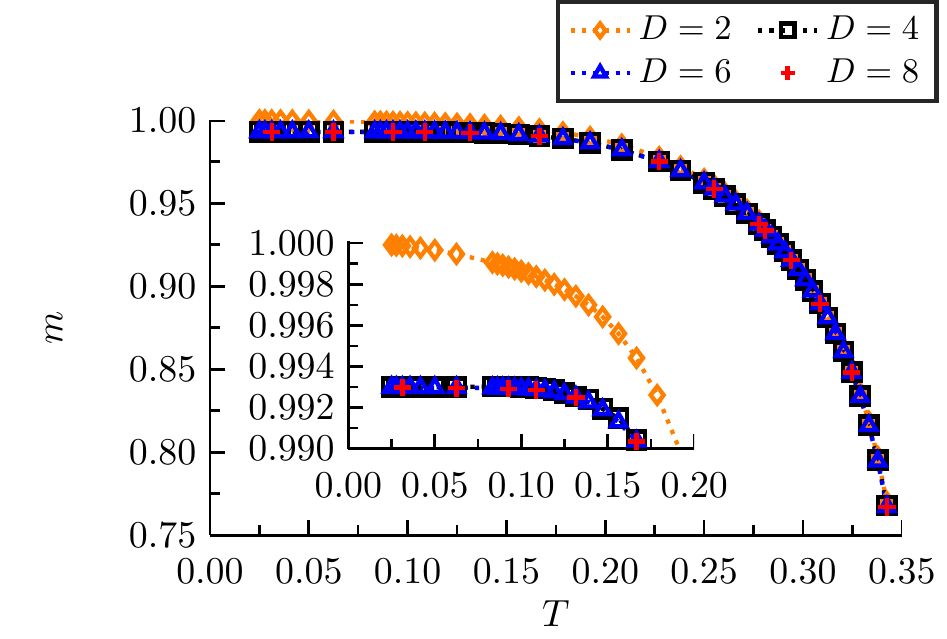}
\vspace{-0.3cm}
\caption{(a)
In (a) the order parameter $m(T)$ (3) as a function of temperature $T$
in the low temperature phase.
The inset (b) shows the zoom on $m(T)$ in the low temperature range 
$T<0.18$. The results demonstrate fast convergence in $D$:
only $D=2$ exhibits a different behavior,
while $D=6$ and $D=8$ data overlap with those for $D=4$.
}
\label{fig:ordlowT}
\end{figure}

In our simulations we use the algorithm described in detail in Ref. 
\cite{Cza16}. In prticular we use corner  matrix renormalization (CMR) 
to contract approximately  tensor networks representing 
thermal states \cite{CMR,CMRPEPS}. To reach  convergence 
of the observables $m$ and $\chi$ approximately
$10$ iterations of the optimization loop were necessary. The
isometries at the beginning of the loop were initialized by a local
truncation scheme based on higher-order singular value decomposition.
The CMR procedure  made $\sim 1000$ 
iterations in the whole loop. The further away from the phase 
transition, the fewer CMR iterations were necessary to reach 
convergence.

Linear susceptibility $\chi(T)$ defined by Eq. (\ref{chiA0}) was 
calculated from a finite difference of the order parameter $\delta m$ 
corresponding to finite difference of the symmetry breaking field 
$\delta h = 10^{-6}$:
\begin{equation}
\chi = \frac{\delta m}{\delta h},
\label{chinum}
\end{equation}
where $\delta m = m(h=\delta h) - m(h=0)$.
Fig.~\ref{fig:susanis} shows that $\chi(T)$ is already converged in 
$\delta h$ for $\delta h=10^{-6}$.
More accurate benchmark of $\delta h$ convergence is given by 
Table~\ref{tab:deltah} showing that decreasing $\delta h$ further 
results in changes of fitted $\gamma$ and $T_c$ that are negligible 
as compared to their dependence on $D$ or the range of $T$.

All simulations were done in {\sc Matlab} with an extensive use of 
the {\sc Ncon} procedure \cite{encon}. To give an idea of the actual 
time and computer resources needed to generate the data, the most 
challenging data points nearest to the phase transition, with the 
largest bond dimensions $D=11$ and $M=72$, required $1-2$ days on 
a desktop.

\section{Simulation of the low temperature phase}
\label{sec:lowT}

The entanglement in the low $T$ phase is small enough to converge the
curve $m(T)$ in $D$ already for $D=4$, see Fig.~\ref{fig:ordlowT}.

Thanks to a short correlation length at low temperature, the 
calculations are much less demanding numerically than close to the 
critical point. Because of that we were  able to generate the data 
shown in Fig.~\ref{fig:ordlowT} during one day using a laptop.

\end{document}